\documentclass[aps,prl,showpacs,twocolumn,superscriptaddress]{revtex4}
\usepackage{graphicx}
\usepackage[latin1]{inputenc}
\usepackage{textcomp}
\usepackage{mathptmx}

\usepackage{color}

\begin{document}

\title{Localization by Dissipative Disorder: a Deterministic Approach to Position Measurements}

\author{Giovanni Barontini}
\email{gbarontini@lkb.ens.fr}
\affiliation{Laboratoire Kastler Brossel, ENS, UPMC-Paris 6, CNRS, 24 rue Lhomond, 75005 Paris, France}

\author{Vera Guarrera}
\email{vera.guarrera@obspm.fr}
\affiliation{LNE-SYRTE, Observatoire de Paris, CNRS, UPMC, 61 avenue de l'Observatoire, 75014 Paris, France
}

\date{\today}

\begin{abstract}
We propose an approach to position measurements based on the hypothesis that the action of a position detector on a quantum system can be effectively described by a dissipative disordered potential. We show that such kind of potential is able, via the dissipation-induced Anderson localization, to contemporary localize the wavefunction of the system and to dissipate information to modes bounded to the detector. By imposing a diabaticity condition we demonstrate that the dissipative dynamics between the modes of the system leads to a localized energy exchange between the detector and the rest of the environment -the "click" of the detector- thus providing a complete deterministic description of a position measurement. We finally numerically demonstrate that our approach is consistent with the Born probability rule .
\end{abstract}


\maketitle 

A quantum system, when measured, is never found in a superposition state. Moreover the quantum algorithm is not able to predict the outcome of the single measurement, only its probability to happen. This constitutes the measurement problem in quantum mechanics. In the Copenhagen interpretation, the problem is circumvented by means of projector operators. The outcome of a single measurement follows from the non-deterministic "collapse" of the wave-packet, i.e., from an elusive process that gives rise to an individual, objective and localized real event: the "click" of the detector \cite{omnes,vonneumann}. In the last decades the decoherence program has provided a more involved point of view on the measurement processes \cite{zurek,raimond1,schloss}. Decoherence and einselection respectively destroy the coherence between the states of a quantum system and select a preferred set of states which are resilient to the environmental action. Nevertheless the decoherence approach has not completely solved the measurement problem. In particular regarding position measurements, it is not able to practically provide a mechanism that leads to a definite outcome in a single measurement, i.e., to the "click" of the detector. This is mainly due to the fact that all position measuring devices,  being them photographic plates, bubble chambers, CCD arrays etc.. are highly complicated and structured systems. Given the huge number of particles and degrees of freedom involved, a complete description of their dynamics is practically impossible and the measurement apparatus is usually described by means of effective models \cite{zurek,schloss,joos,mon,caldeira,brownian} or effective potentials \cite{allcock,muga2}. As a consequence, the relevant mechanisms responsible for the measurement outcomes are difficult to be pointed out. In the vast majority of the above mentioned models, such mechanisms are hidden in the interaction with a generic "noisy" bath. Interestingly, specifically concerning position measurements, to the best of our knowledge the spatial disorder on the measurement device has never been considered as possible responsible for the localization of the particle under detection. The hypothesis that a position measurement device can be effectively described by a dissipative disordered potential appears natural when dealing with non-idealized detector. First, at the microscopic scale, Nature always tends to arrange in a disordered way. Moreover, even in the case the detector was consisting of the most perfect crystal, on its surface the dynamical chemical equilibrium with the surrounding gas and the presence of already detected particles strongly modify the potential, effectively disordering it. Here we thus consider a position measurement device as a disordered purely absorptive potential \cite{allcock,hist,rus}. We demonstrate that a disordered absorptive potential is indeed able to reduce an initially delocalized wavefunction to a localized one via a particular form of einselection: the dissipation-induced Anderson localization. Localization by the environment has be achieved in a large number of models, see e.g. \cite{joos,caldeira,hist,rus,loc,loc2}, but Anderson-like mechanisms have so far never been considered. In addition here we show that, making use of a diabaticity condition, the absorption process induces an energy exchange between the detector and the rest of the environment which is localized in space and which represents the measurement outcome, namely the "click" of the detector. Given a certain wavefunction for the quantum system, the actual realization of the potential deterministically sets the position of localization and the outcome of the single measurement. The practically unpredictable modification of the disordered potential from one measurement to another changes such position, making the output essentially aleatory. The repetition of a suitable high number of such kind of measurement processes allows to reconstruct the density distribution of the quantum system, in accordance with the Born probability rule. In contrast with spontaneous collapse theories \cite{pearl, gir,proj}, our approach does not require any modification of the Schr\"odinger equation or the introduction of underlying noise fields. 

\begin{figure}
\begin{center}
\includegraphics[width=0.45\textwidth]{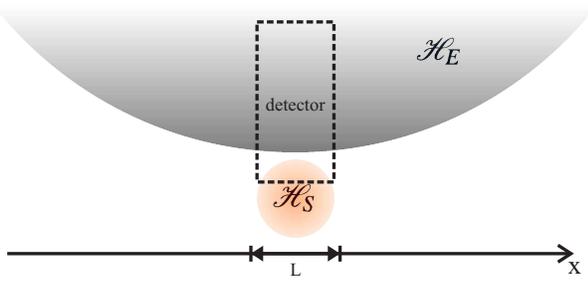}
\end{center}
\caption{(Color online) Pictorial representation of a position measurement. A generic quantum system, whose associated Hilbert space is $\mathcal{H}_S$ is confined but delocalized in a region of space of length $L$. The environment, with Hilbert space $\mathcal{H}_E$, is instead spread all over the space. At $t=0$ the environment and the quantum system interact through a detector, whose spatial extension is at least $L$. The vertical dimension is used to pictorially represent the extension of the Hilbert spaces.} 
\end{figure}

We start our treatment considering the idealized situation of a particle and the surrounding environment and defining the corresponding Hilbert spaces as $\mathcal{H}_S$ and $\mathcal{H}_E$. The total Hilbert space can be written as the tensor product $\mathcal{H}=\mathcal{H}_S\otimes\mathcal{H}_E$ and the states can be written as $\Psi=\Psi_S\Psi_E$, obeying the Schr\"odinger equation $H\Psi=E\Psi$. We additionally impose the normalization condition $\int|\Psi_S|^2=1 \ \forall \ t$. Let us now consider the subspace $\mathcal{H}_D$ and its projector $Q$, so that $Q\mathcal{H}=\mathcal{H}_D$, which are associated with a position detector (see Fig.1). The detector is where the environment and the particle interact and we suppose that for $t\leq0$ such interaction is off and $Q\Psi_S(t<0)=0$.   By means of the Feshbach partition theorem \cite{feshbach,zwanzig,muga} it is possible to demonstrate that the wavefunction outside the detector $\Psi_O=P\Psi=(1-Q)\Psi$ evolves under the action of both Hermitian and non-Hermitian terms. By formally tracing out the environment degrees of freedom and making use of a Markovian approximation for the detector configurations \cite{muga,lude,nota1}, we can obtain a simple expression for the evolution of the wave vector $\psi$ of the particle outside the detector:
\begin{equation}
i\hbar\partial_t\psi=H_S\psi+D_{int}\psi=H_S\psi-i\theta(t-0)V(x)\psi,
\end{equation}
where the theta-function renders the sudden switch-on of the detector at $t=0$. Similar approaches have been used for example in \cite{hist,rus,allcock, muga2, steck}. Irrespectively on the details of the microscopic interactions, being them chemical reactions, ionization mechanisms etc.., our ansatz is that $V(x)$ has a disordered spatial dependence \cite{joos,caldeira}. Equation (1) describes the evolution of the particle as an open system connected to the detector, where the non unitary evolution implies that the norm of $\psi$ is not conserved, i.e., $d_t\left[\int|\psi|^2\right]\neq0$. It then holds that $\Psi_S=\psi+\phi$ with $\Psi_S(t\leq0)\equiv\psi(t\leq0)$ and $d_t\left[\int|\Psi_S|^2\right]=d_t\left[\int|\psi+\phi|^2\right]=0$.
The two components $\psi$ and $\phi$ are not necessarily orthogonal. In fact, owing to the elimination of the environment degrees of freedom, it is not possible to define a projector $P_S$ (corresponding to $P$) such that $\psi = P_S\Psi_S$ (so that $\phi\neq(1-P_S)\Psi_S$). The interpretation of the two wavefunctions is readily done: $\psi$ is the component which is external to the detector while $\phi$ corresponds to the internal component, e.g. to bound states inside the detector. The integrals of their density distributions correspond respectively to the survival probability in the initial system and to the probability of the particle to be absorbed by the detector and they are obviously bounded to one, i.e., $\int|\psi|^2\leq1$ and $\int|\phi|^2\leq1$. The equation for the evolution of $\phi$ could be obtained with a similar procedure starting from $\Psi_D$. Nonetheless, and this is the second ansatz of our treatment, this is not needed. Indeed, given eq. (1), the complete evolution of the two components of $\Psi_S$ during the measurement process is already known if we impose the diabaticity condition 
\begin{equation}
d_t[\Psi_S]_{t<t_M}=0,
\end{equation} 
where $t_M$ is the time at which the detector clicks. This condition follows from the fact that here we are interested in strong measurements, i.e., in measurements where the detector is able to rapidly and irreversibly localize the particle. For $t<0$, when the environment and the system do not interact, it holds $H\Psi_S=E_S\Psi_S$ with $E_S\ll E$. The time evolution in the system then is $\Psi_S(t)=\exp[-iE_St/\hbar]\Psi_S(\tau)$, where $\tau<0$ sets the time at which the quantum system was prepared.  We can define the measurement time as $t_M\simeq\hbar/\Im{(E_{int})}$, with $D_{int}\chi=E_{int}\chi$. The absorption term $D_{int}$ for our projective detector must then satisfy the condition $\Im{(E_{int})}\gg E_S$. From which follows that $\Psi_S(t_M)=\exp[-iE_S/\Im{(E_{int})}]\Psi_S(0)\simeq\Psi_S(0)$. The diabaticity condition can thus also be written as $\langle |H_S|\rangle\ll\langle |D_{int}|\rangle$. Similar conditions are found in the spontaneous collapse theories \cite{proj}.

\begin{figure}
\begin{center}
\includegraphics[width=0.45\textwidth]{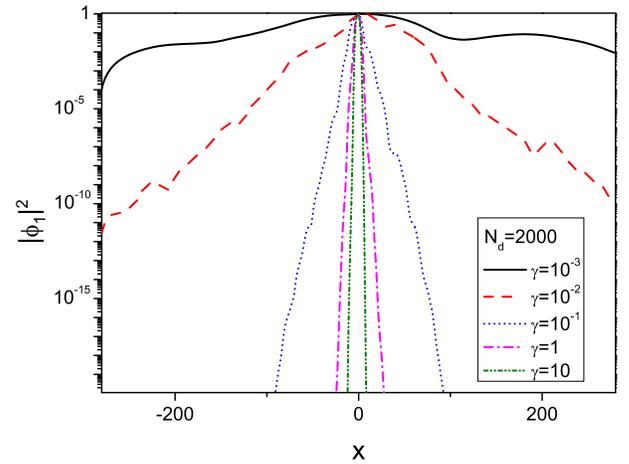}
\end{center}
\caption{(Color online) Typical normalized density distribution of the eigenstates $\phi_1$ for the disordered dissipative potential described in the text, for different amplitudes of the speckles. The vertical logarithmic scale allows to appreciate the exponentially localized character when $\gamma>10^{-3}$.} 
\end{figure}

\begin{figure*}
\begin{center}
\includegraphics[width=0.45\textwidth]{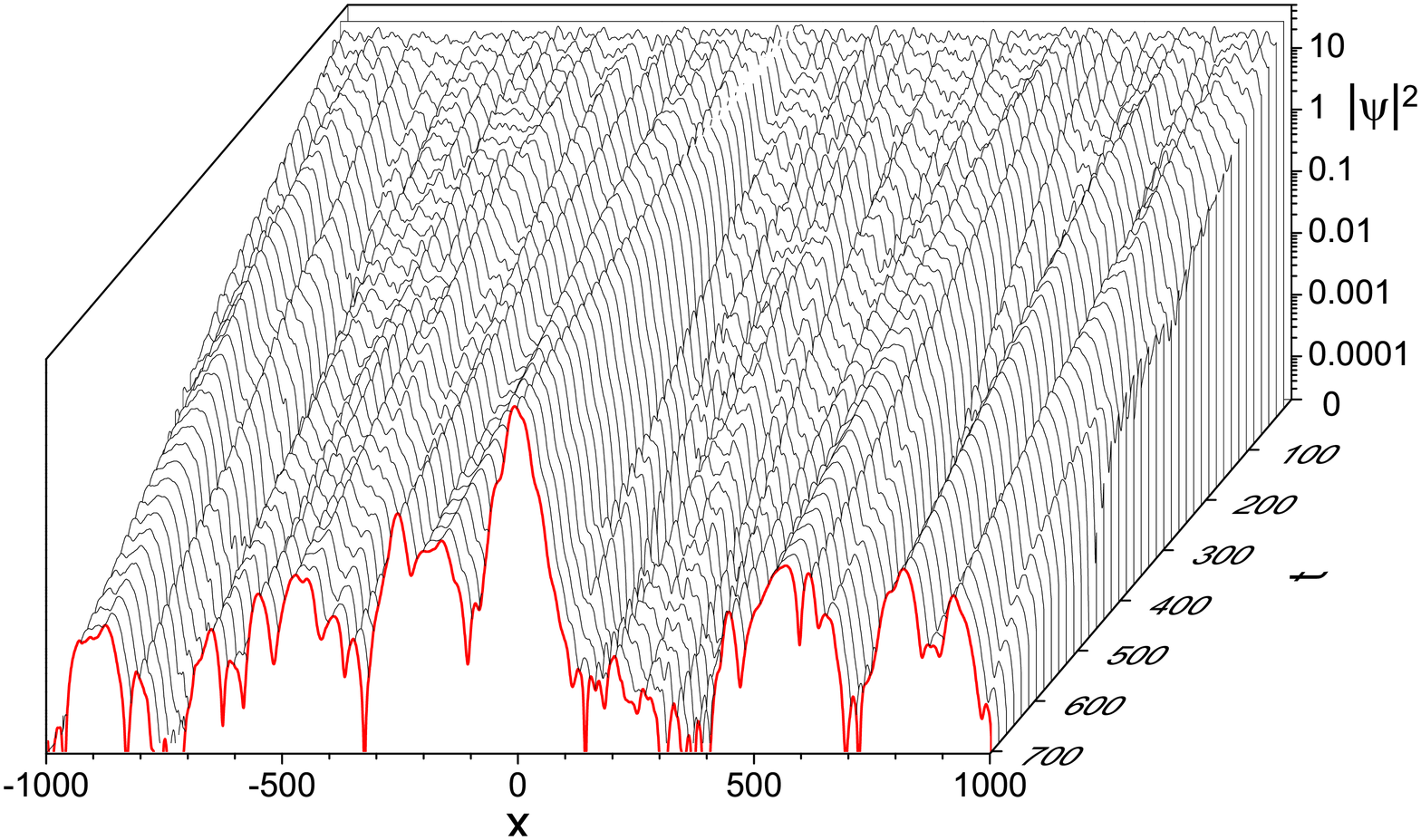}
\includegraphics[width=0.45\textwidth]{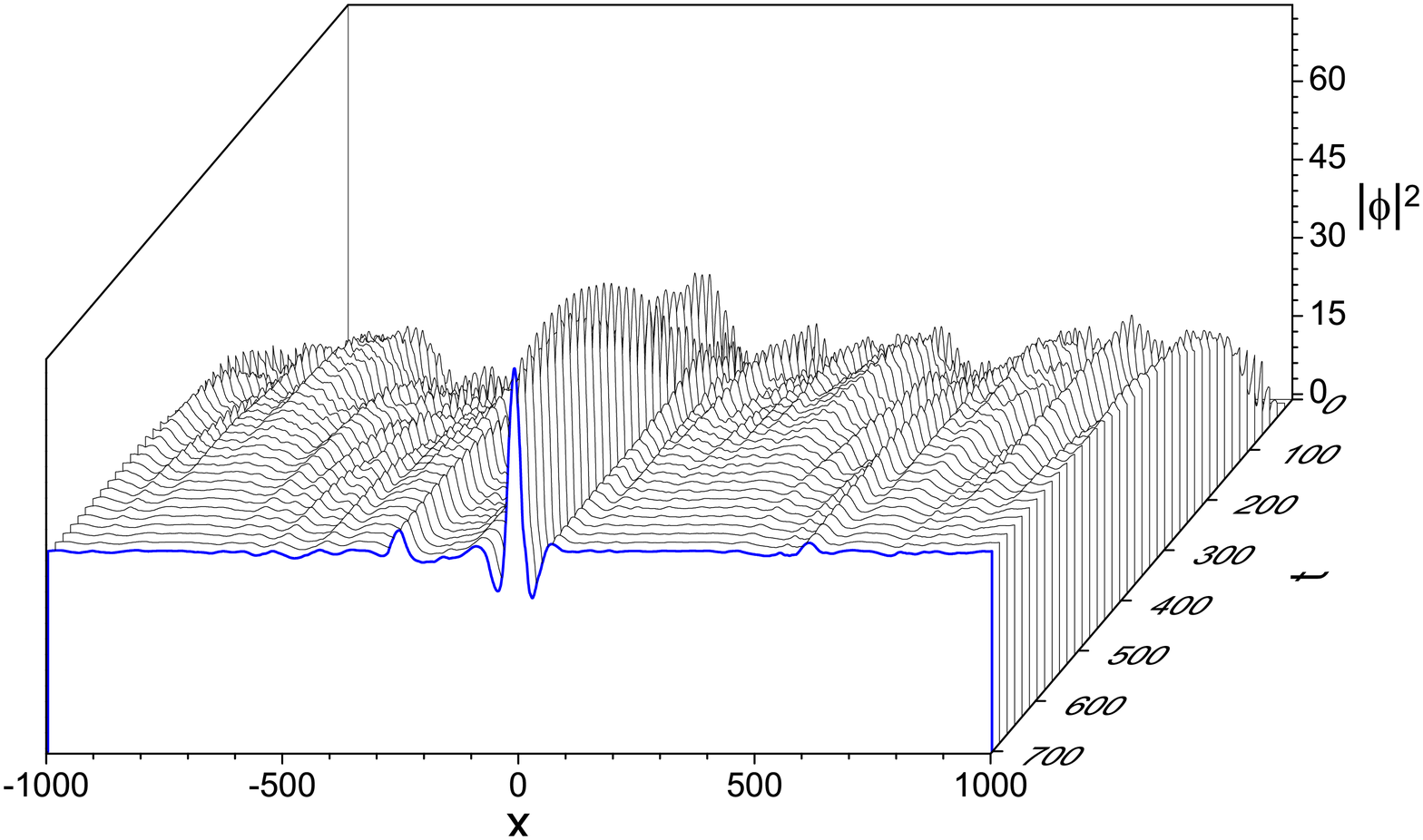}
\end{center}
\begin{center}
\includegraphics[width=0.45\textwidth]{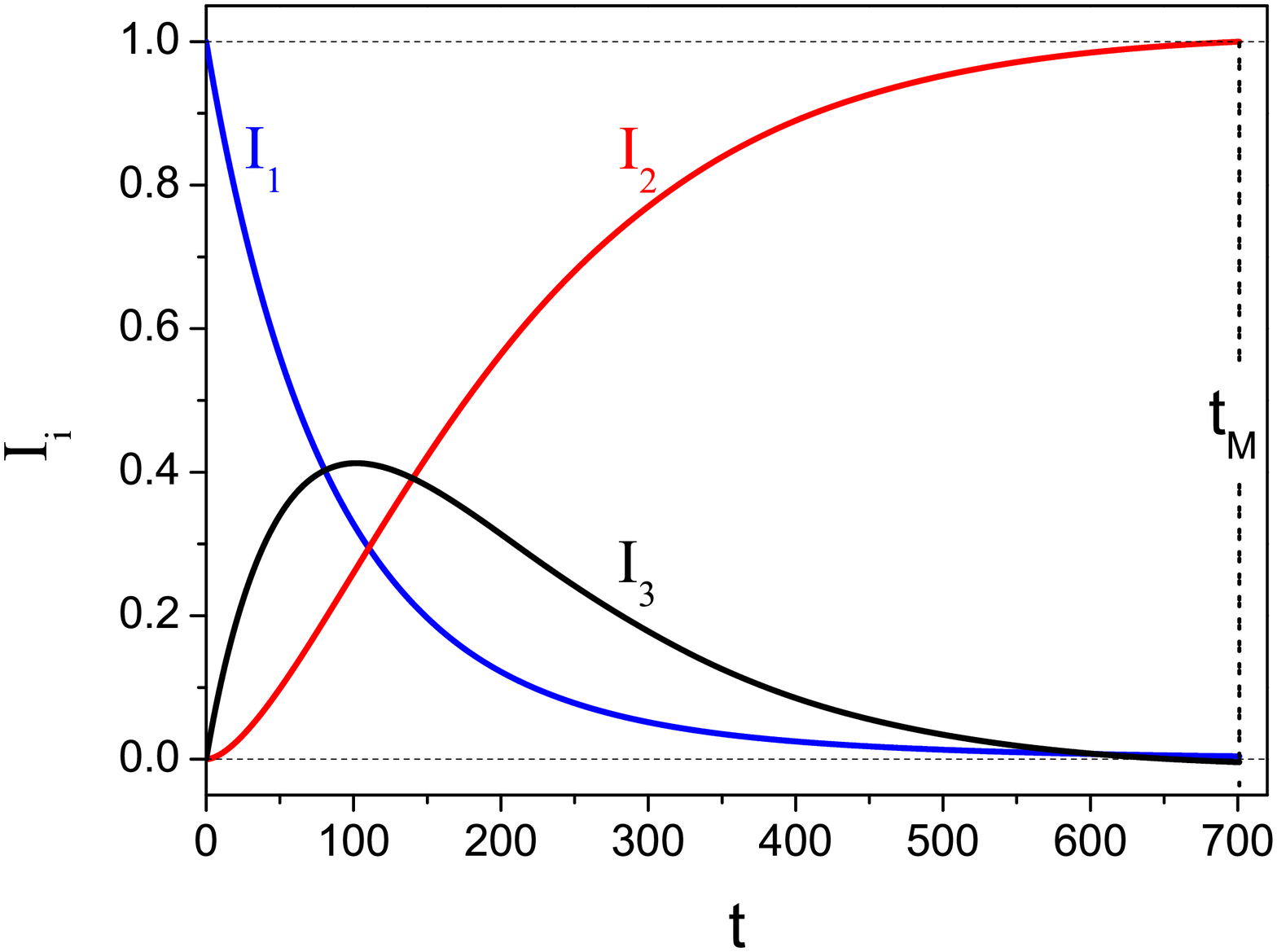}
\includegraphics[width=0.45\textwidth]{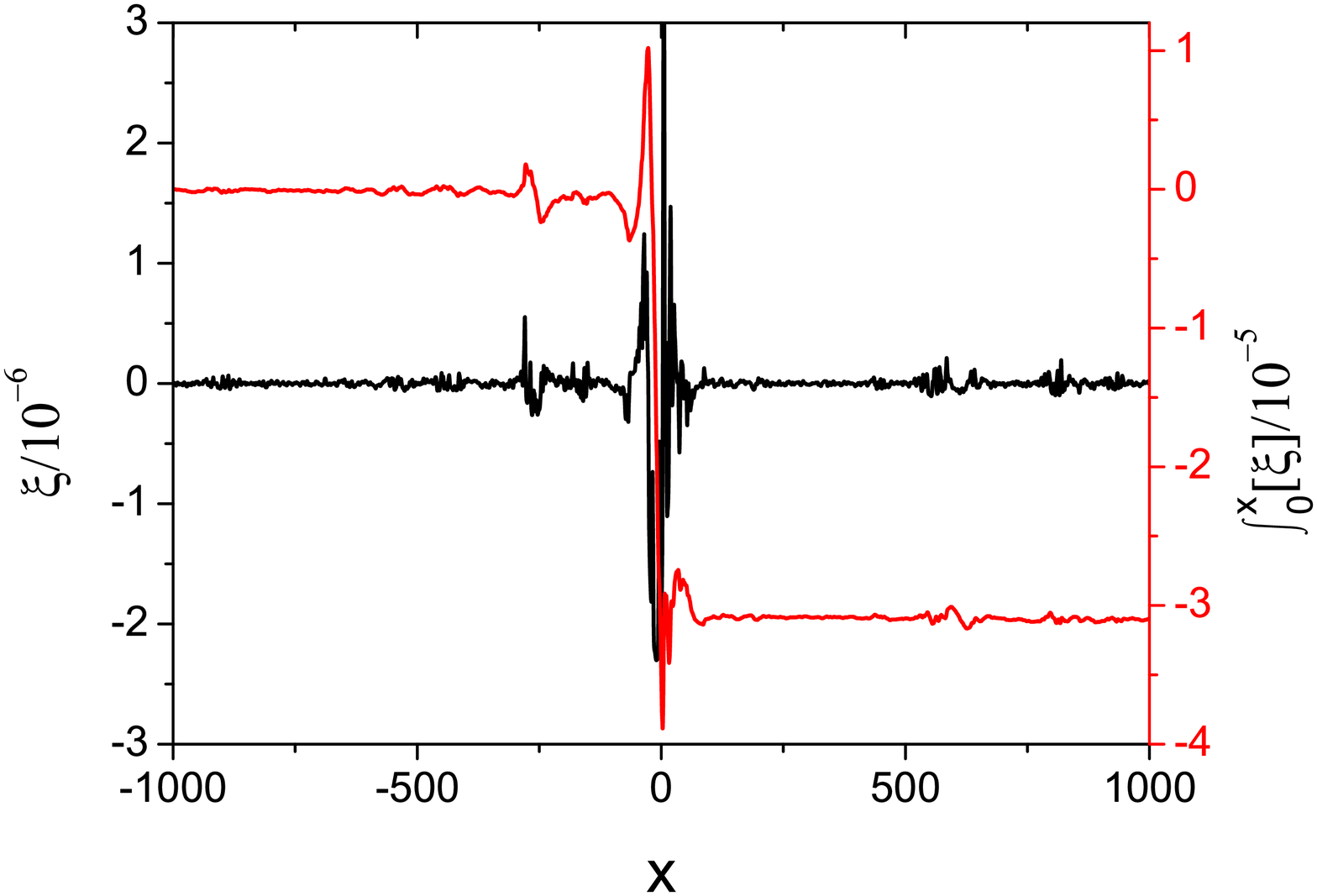}
\end{center}
\caption{(Color online) Dynamics of the dissipation-induced Anderson localization. a) and b) The evolution of $|\psi(t)|^2$ and of $|\phi(t)|^2$ as a function of time until the measurement time $t_M$. The vertical scale in a) is logarithmic highlighting the exponential localization of the final state.  c) $I_1$, $I_2$ and $I_3$ as a function of time. d) The energy density $\xi$ and its integral $\int_0^x[\xi]$ as a function of $x$ for $t=t_M$.} 
\end{figure*} 

Given the above scenario we now describe the process that, starting from eq. (1), distillates the states that are more resilient to the dissipative action. We consider a finite size system of length $L$ (corresponding to the confinement volume of the particle or to the detector size), whose discrete spectrum is characterized by eigenvalues  $E^{tot}_n$ and eigenvectors $\phi_n(x)=\langle x|n\rangle$ with $E^{tot}_n=E_n+iV_n$. Given a certain initial wavefunction $\langle x|\psi(0)\rangle$ one obtains that its evolution can be expressed in term of the eigenstates $|n\rangle$ as \cite{moratti}:
\begin{eqnarray}
\langle x\vert \psi(t) \rangle &=&\sum_n \langle x \vert n \rangle \langle n'\vert \psi(0)\rangle e^{-iE^{tot}_{n}t/\hbar}\nonumber\\
&=&\sum_n c'_n \phi_n(x)e^{-i(E_n+iV_n)t/\hbar}
\end{eqnarray}
where $c_{n'}=\langle n' \vert \psi(0) \rangle$. The right and left eigenstates are
complex conjugate $|n'\rangle=|n\rangle^*$, from which it follows that
$c_{n'} = |c_n|e^{-i\theta_n}$ , being $c_n = \langle n|\psi(0)\rangle = |c_n|e^{i\theta_n}$ the coefficient
of the expansion over the right eigenstates. For the density one obtains:
\begin{eqnarray}
&& |\psi(x,t)|^2=\sum_n \vert c_n\vert^2 \vert \phi_n (x)\vert^2 e^{2\frac{V_nt}{\hbar}}+  \nonumber\\
&& \sum_{n\neq m} \vert c_n\vert \vert c_{m}\vert \phi^{*}_{n}(x)\phi_{m}(x)e^{i \Delta \varphi_{nm}(t)}e^{\frac{(V_n+V_{m})t}{\hbar}}
\end{eqnarray} 
with $\Delta \varphi_{nm}(t)=(\theta_{m}-\frac{E_{m}t}{\hbar})-(\theta_{n}-\frac{E_{n}t}{\hbar})$. Since $V_n<0$ we have that \emph{i)} in the first sum, only the terms with smaller $|V_n|$ survive
at long times and \emph{ii)} in the second sum, at long times only the terms with smaller $|V_n|$ and $|V_{m}|$ survive as well. In particular, since $|V_n|<|V_{m>n}|$ the terms of the second sum are suppressed faster than the first sum. The total effect is thus a particular form of einselection: the dissipative action of the detector irreversibly distillates the sets of states with the smaller imaginary energy $V_n$.

As an example of dissipative disorder potential we choose $V(x)=\sum_{j=1}^{N_d}\gamma\exp(-(x-x_j)^2/(2\sigma^2))$, hence a speckle-like potential. The centres of the speckles $x_j$ are uniformly randomly distributed over the detector size $L$. $N_d$, $\gamma$ and $\sigma$ are respectively the number of speckles, the amplitude and the size of the speckles. Moreover we set $H_S = \hbar^2\nabla^2/(2m)$. In the following we use $\sigma$ as length unit and then $\hbar^2/(2m\sigma^2)$ as energy unit, being $m$ the mass of the particle. In Fig. 2 we show the density distribution of the states with the smallest $V_n$, for a given realization of the disorder potential with $L=2000$, $N_d=1000$ and different values of $\gamma$. In case the dissipation is sufficiently strong they are exponentially localized so we call the distillation process \emph{dissipation-induced Anderson localization} \cite{anderson, footnote2}. We remark that, while Anderson localization originates from the destructive interference of waves in disordered conservative potentials, here the localization results from the selection of the states which are most resilient to a random distribution of absorbing potentials.

Hence eq. (1) with a disordered potential $V(x)$ leads to the contemporary localization and dissipation of $\psi$ while, given eq. (2), the time evolution of $\phi$ is then simply set by the equation $d_t\phi=-d_t\psi$. The aim of the measurement protocol is then to obtain $|\psi(x,0)|^2 = |\Psi_S(x,0)|^2$. From the normalization condition it follows that $\int|\psi|^2+\int|\phi|^2+2\Re\int(\psi^*\phi)=I_1+I_2+I_3=1 \ \forall \ t$.  At $t=0$, when the measurement starts, the detector entangles $\psi$ and $\phi$, indeed $I_3\neq0$. Moreover it starts to acquire information dissipating it from $\psi$ and transferring it to $\phi$.  The "click" of the detector happens when $I_2 = 1$, i.e., when the available information has been fully absorbed. In this case the detector has the certainty about the particle existence and additionally it cannot further absorb ($I_2\leq1$), hence the measuring process must stop. The diabaticity condition allows such scenario to manifest at a finite time $t=t_M$. Indeed it makes possible to reach the condition $I_2(t_M)=1$ with $I_1(t_M)=-I_3(t_M)$. Stopping the measurement implies the  disentanglement of $\phi$ and $\psi$ which, in turn, implies the orthogonality of the two wavefunctions. Moreover, since $I_2=1$, it also implies that $\psi(t>t_M)=0$. Notably, in contrast with spontaneous localization theories, our treatment does not suffer from tails problems \cite{schloss, tail}.

Ending the measurement thus requires an energy amount to disentangle $\psi$ and $\phi$ and to annihilate $\psi$. The environment, which continuously interacts with the detector, can provide such energy, corresponding to the "click" of the detector. At the beginning of the measurement the energy of the particle is $E_S=\int[\psi^*(0)H\psi(0)]$ while immediately after the "click" it becomes $E_{S'}=\int[\phi^*(t_M)H\phi(t_M)]>E_S$. The difference between $E_S$ and $E_{S'}$ is mainly \cite{footnote} the energy associated with the entanglement at the end of the measurement $\mathcal{E}=\int[\phi^*(t_M)H\psi(t_M)+\psi^*(t_M)H\phi(t_M)]=\int[\xi(t_M,x)]$, being $\xi$ the energy density. By ending the measurement process  such energy is indeed absorbed by the detector from the environment and, in case it is localized in space, represents the read out of the measurement. This can happen if at least one of the two wavefunctions is localized before the "click", condition ensured in our case by the dissipation-induced Anderson localization of $\psi$. 
The collapse in the frame of our treatment is thus a consequence of the disentanglement, which implies $\psi(t\geq t_M)=0$. Formally we can include the "click" and the end of the measurement process as:
\begin{equation}
\Psi_S(t)=\theta(I_2(t)-1)\psi(t)+\phi(t).
\end{equation}
Immediately after the measurement the particle has been fully absorbed by the detector and it is still completely described by the wavefunction $\phi$ (with unitary norm). 

\begin{figure}
\begin{center}
\includegraphics[width=0.47\textwidth]{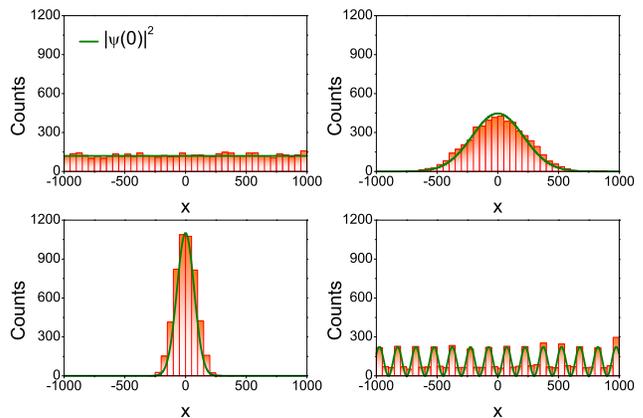}
\end{center}
\caption{(Color online) Histograms obtained after 5000 simulations
of position measurements following our algorithm. In each panel the
results are reported together with the corresponding initial density distribution $|\psi(0)|^2$ (line). The bin size is set to be safely larger than the typical FWHM of $\xi$.} 
\end{figure}

In order to verify our approach we have performed a series of numerical simulations choosing the dissipative potential $V(x)$ to have the form described above. As shown in Fig. 2 the states with the smallest imaginary parts of the spectrum are exponentially localized. In Fig. 3 we report an example of the dynamics, with $L = 2000$, $N_d= 1000$ and $\gamma = 10^{-2}$ starting from $\psi(0)$ being completely delocalized over $L$. Such choice of $\gamma$ and $N_d$ fulfil the condition $\langle |H_S|\rangle\ll\langle |D_{int}|\rangle$. In Fig. 3a and 3b we show the behaviours of $|\psi(x)|^2$ and $|\phi(x)|^2$ during the dynamics. In particular in Fig. 3a it is possible to appreciate the exponential localization of $|\psi(x)|^2$ accompanied by the dissipation. The action of the dissipative potential gradually suppresses the norm of $\psi$ ($I_1$) while, what is dissipated from $\psi$ represents a gain for $\phi$, whose norm $I_2$ goes from 0 to 1 at $t=t_M$ (Fig. 3c). The cross-term $I_3$, which takes into account the overlap of $\psi$ and $\phi$, starts from 0, has its maximum when $I_1\simeq I_2$ and then goes below 0 around $t_M$.  In Fig. 3d we plot the energy density $\xi(x)$ at $t=t_M$, which is localized (as also evident from its integral) approximately in the same position as $|\psi(t_M,x)|^2$. The width of $\xi$ naturally sets the resolution of the detector. For the particular disordered potential we have chosen, we have found that the typical FWHM of $\xi$ is $\simeq30$. 

Our approach to a position measurement therefore allows a localized energy exchange starting from a delocalized wavefunction. We now demonstrate that our approach is compatible with the Born probability rule, i.e., that is possible to reconstruct the space dependence of modulus squared of an arbitrary initial wavefunction $|\psi(t = 0,x)|^2$. As in standard experiments this can indeed be inferred summing up the results of a suitable number of single position measurements (ideally infinite). We start from four different kinds of initial wavefunctions $\psi(t = 0,x)$: a completely delocalized one
(flat), two gaussians with different widths and a sine. For this set of simulations we use the same kind of speckle potential described above. We let the system evolve according to eq. (1), (2) and (5) for 5000 different realization of the disordered potential. In a few cases ($<1\%$) the protocol is not able to localize the particle in the maximum time allowed and such runs are discarded. In the other cases, when $I_2 = 1$ the dynamics is stopped and $\xi(x)$ is calculated. We then build the histogram of the positions where $\xi(x)$ is localized with a binning of 50, safely larger than the average FWHM of $\xi$ itself. In Fig. 4 we report the results of the simulations together with the corresponding initial states for all the four cases; the agreement is excellent demonstrating the plausibility of our approach.

In summary, we have introduced an approach to position measurements mainly built on the hypothesis that a position detector is a disordered absorptive potential. We have shown that the interaction of such detector with the quantum system gives rise to the dissipation-induced Anderson localization. The subsequent "click" of the detector is obtained by imposing a diabaticity condition. We have performed a numerical experiment following the above described measurement process and we have been able to excellently reconstruct four different density distributions. The onset of the dissipation-induced Anderson localization could be observed in controlled systems like Bose-Einstein condensates under the action of suitably engineered dissipative potentials \cite{us}. In such experiments the competition of disorder and order in the localization of a particle under measurement could indeed be tested. Our approach could also be compared with other models of wavefunction reduction \cite{proj}, chaotic systems \cite{reizen} or bohmian mechanics \cite{bohm}. Additionally its extension to other contexts, like to the quantal definition of the time of arrival \cite{allcock, mugatime} may lead to interesting new insights.

\paragraph*{Acknowledgements}
We are grateful to M. Modugno for reading the manuscript and for fruitful discussions.


\begin{thebibliography}{}
\bibitem{nota1}
The Markovian approximation implies that the absorptive potential does not change during the single measurement and that two successive configurations do not depend the one on the other.
\bibitem{footnote2}
Note that starting from a certain time $t_1$, which depends on the details of the potential, only the lowest state survives.  Such a state, once distilled, remains unaffected by the dissipative potential, only its amplitude is constantly reduced.
\end{thebibliography}
\end{document}